\documentclass[showpacs,prl,onecolumn,aps,superscriptaddress,preprintnumbers,nofootinbib]{revtex4}
\usepackage[T1]{fontenc}
\usepackage[latin9]{inputenc}
\usepackage{amsmath,amssymb}
\usepackage{epsfig}
\usepackage{bm} 
\usepackage{graphicx}
\usepackage{mathrsfs}
\usepackage{amsmath}
\usepackage{amsfonts}
\usepackage{epstopdf}
\usepackage{color}
\usepackage{pifont}
\usepackage{relsize}
\usepackage{mathtools}
\def\slashchar#1{\setbox0=\hbox{$#1$}     		
   \dimen0=\wd0                                 	
   \setbox1=\hbox{/} \dimen1=\wd1               	
   \ifdim\dimen0>\dimen1                        	
      \rlap{\hbox to \dimen0{\hfil/\hfil}}      	
      #1                                        	
   \else                                        	
      \rlap{\hbox to \dimen1{\hfil$#1$\hfil}}   	
      /                                         	
   \fi}

\renewcommand{\vec}{\boldsymbol}
\newcommand{\beq}{\begin{equation}}
\newcommand{\eeq}{\end{equation}}
\newcommand{\bea}{\begin{eqnarray}}
\newcommand{\eea}{\end{eqnarray}}
\newcommand{\ba}{\begin{array}}
\newcommand{\ea}{\end{array}}

\def\eq#1{{Eq.~(\ref{#1})}}
\def\fig#1{{Fig.~\ref{#1}}}
\newcommand{\bas}{\bar{\alpha}_S}
\newcommand{\as}{\alpha_S}
\newcommand{\nn}{\nonumber}
\newcommand{\bg}{ \bar{\gamma}}

\newcommand{\Lb}{\left(}
\newcommand{\Rb}{\right)}
\newcommand{\h}{\frac{1}{2}}

\newcommand{\pom}{I\!\!P}

\newcommand{\Y}{\tilde Y}
\newcommand{\te}{\tilde \eta}\newcommand{\intl}{\int\limits}
\begin{document}

\title{ Dipole-dipole scattering at high energy in the Pomeron field theory with Braun Hamiltonian and beyond  }

\author{Eugene Levin}
\email{leving@tauex.tau.ac.il}
\affiliation{Department of Particle Physics, School of Physics and Astronomy,
 Faculty of Exact Science, Tel Aviv University, Tel Aviv, 69978, Israel}
\date{\today}


\begin{abstract}
 In this paper we find that the scattering matrix for dipole-dipole interaction in the saturation region has the form: ${\cal S}^d_d \xrightarrow{z \,\gg\,1} \exp\Lb - C^d_d z^2\Rb \propto \Lb {\cal S}_{BK}\Rb^4 = \exp\Lb - 4\,C_{BK} z^2\Rb$, where $z = \bas \kappa Y\,+\,\ln\Lb \frac{r^2}{r'^2}\Rb$ for interaction of a dipole $r$ at rapidity $Y$ with dipole $r'$ at  rest. $S_{BK}$ is the S-matrix for  the  Balitsky-Kovchegov amplitude.  All constants are determined in the text. The proof is given in the same theoretical  framework for both cases (${\cal S}^d_d$ and ${\cal S}_{BK}$): the Pomeron interaction  which takes into account the Pomeron vertices in the leading  $1/N_c$ approximation ($N_c$ is the number of colours). This result  is in  striking contradiction with 'rare' fluctuation approach as well as with summing the large Pomeron loops. These lead to  ${\cal S}^d_d \propto \sqrt{{\cal S}_{BK}}$ at large $z$. In the paper we 
 collect  arguments supporting the idea that the sum  of large  Pomeron loops can be trusted  in a wide region of energy: $Y \leq 1/\as^4$.  In addition we discuss the influence  of the structure of the Hamiltonian on the asymptotic behaviour of the scattering amplitudes using the exactly solvable one dimensional model as our theoretical ground.

 \end{abstract}
\maketitle

\vspace{-0.5cm}
\tableofcontents

\section{Introduction}
 The BFKL\footnote{BFKL stands for Balitsky, Fadin, Kuraev and Lipatov\cite{BFKL}.} Pomeron calculus is the effective theory for high energy interaction in QCD. Its main ideas are based on 
 the reggeon approach of Ref.\cite{Gribov} and have been developed over the years in the framework of QCD  \cite{BFKL,GLR,GLR1,MUQI,MUDI,LIREV,LIP,LipatovFT,BART,BKP,MV,MUSA,KOLE,BRN,BRAUN,B,K,AKLL,ACKLLS,KLL}. It has been shown \cite{BRAUN,KOLU} that the Pomeron calculus describes the Color Glass Condensate (CGC) approach\cite{JIMWLK1,JIMWLK2, JIMWLK3,JIMWLK4, JIMWLK5,JIMWLK6,JIMWLK7,JIMWLK8} (see also Ref.\cite{KOLEB} for a review). 
 
 In Ref.\cite{BRAUN} a Hamiltonian was suggested for the Pomeron interaction which takes into account the Pomeron vertices in the leading  $1/N_c$ approximation ($N_c$ is the number of colours). The main topic that we consider in this paper is to find the asymptotic behaviour of the scattering amplitude for dipole-dipole interaction at high energies. This paper is the direct continuation of our ten-year-old paper \cite{KLL}, here  we complete the estimates of the scattering amplitude at high energies in the vicinities of the fixed points of the equations of motion. The main  result  of this  paper is that, to our great surprise, deep in the saturation region the scattering amplitude has the form:
 \beq \label{I0}
 N^d_d\Lb z\Rb \,\,=\,\,1\,\,-\,\,\exp\Lb - 2 \frac{z^2}{\kappa}\Rb
 \eeq
 where $z\,\, = \,\,\bas \chi\Lb \gamma_{cr}\Rb \,\eta + \xi_{r,r'}$ with $\kappa \equiv  \chi\Lb \gamma_{cr}\Rb $ (see \eq{GACR}) and $ \xi_{r,r'}$ for scattering dipoles $r$ and $r'$ will be discussed in the next section. Comparing this behaviour with the solution\cite{LETU} for the dipole-nucleus scattering amplitude, which is  described by the Balitsky-Kovchegov nonlinear equation \cite{B,K} :
 \beq \label{I01}
 N^d_A\Lb z\Rb \,\,=\,\,1\,\,-\,\,\exp\Lb -  \frac{z^2}{2\,\kappa}\Rb
 \eeq 
 one can see the striking difference. The situation looks even more dramatic if we compare with the prediction for  dipole-dipole scattering from the 'rare' fluctuation approach\cite{IAMU,IAMU1} and from summing  large Pomeron loops\cite{LELT,LE1,LEDIDI,LEDIA,LEAA}:
  \beq \label{I02}
 N^d_d\Lb z\Rb \,\,=\,\,1\,\,-\,\,\exp\Lb -  \frac{z^2}{4\,\kappa}\Rb
 \eeq 
 
 Frankly speaking, we have no reliable explanation why we have such different predictions.
 Summing  Pomeron loops
 has been one of the  difficult problems in the Color Glass Condensate (CGC) approach, without solving which  we cannot consider the dilute-dilute and dense-dense  collisions  of dipole densities.
However, in spite of intensive work 
\cite{MUSA,BRN,BRAUN,AKLL,ACKLLS,KOLU,KOLEB,KLL,KLLL1,KLLL,KLLN,LETU,LELU,KO1,LE11,RS,KLremark2,SHXI,KOLEV,nestor,LEPRI,LMM,LEM,IAMU,IAMU1,KOLU11,KOLUD,BA05,SMITH,KLW,kl},  this problem  has  not been solved. The only progress has been achieved in summing  large Pomeron loops\cite{LELT,LE1,LEDIDI,LEDIA,LEAA}. 
  
 Certainly, summing  the large Pomeron loop could be trusted only in a limited range of energy. We will discuss this topic in this paper. As far as 'rare' fluctuations are concerned, we do not have a reasonable explanation for  both questions: why the 'rare' fluctuation give the same result as summing the large Pomeron loops and why this approach is not able to describe the high energy asymptotic behaviour of the scattering amplitude for dipole-dipole scattering.
 
 However,  the way out could be related to the general  problem of Pomeron interaction: the Braun Hamiltonian\cite{BRAUN}) violates  s-channel unitarity\cite{KLL,KLLL1,KLLL}.  
The situation is even more disturbing, since this intensive work has not led to a  generalization of the Braun Hamiltonian which satisfies  s-channel unitarity.  On the other hand,  \eq{I0} has been proven on the same footing as \eq{I01}, since the Balitsky-Kovchegov Hamiltonian\cite{B,K} has the same difficulties with  s-channel unitarity as the Braun one\cite{KLL,KLLL1,KLLL}.

 The next section is a brief review of  
the BFKL Pomeron interaction in leading order of $N_c$. This interaction can be described by the Braun Hamiltonian\cite{BRAUN} and we discuss the equations of motion in the form of Ref.\cite{KLL}.  In section III we consider the behaviour of the solutions to the equations of motion in the vicinities of the fixed points. We calculate the scattering amplitudes in these kinematic regions  and obtain \eq{I0} for them. 
 Section IV is devoted to summing  large Pomeron loops. We state that our procedure of summing large Pomeron loops satisfies both $s$- and $t$-channel unitarity and can be trusted for $Y\,\leq \,1/\as^4$. This claim is based on 
the one dimensional model of Ref.\cite{MUSA} which satisfies $t$- and $s$- channel unitarity. We show that the sum of large Pomeron loops in this model describes the main features of the high energy asymptotic behaviour of the scattering amplitude. The difference between the sum of large Pomeron loops and the exact scattering amplitude stems from the intercepts of multi Pomeon exchanges which originate  from the contribution of the small Pomeron loops of  size: $\delta Y \sim \Delta$. In the Conclusions we summarize  the results.

\begin{figure}
\centerline{\epsfig{file=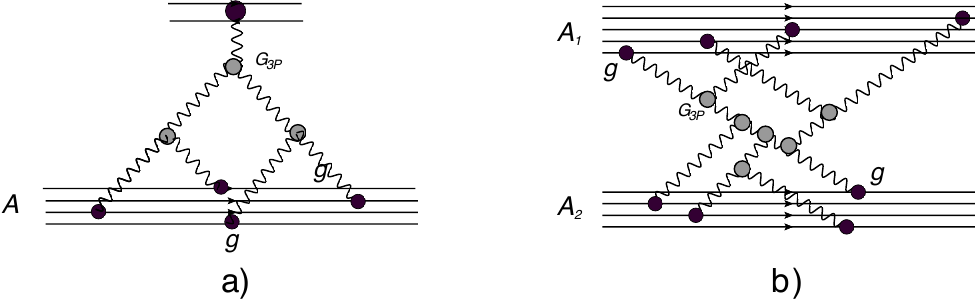,width=140mm}}
\caption{The Pomeron  diagrams that are described by the equation of motion. The 'fan'  diagrams of  \fig{netset}-a 
correspond to the BK equation, which is the equation of motion for the BK Hamiltonian. The net diagrams of \fig{netset}-b appear from the equation of motion for the Braun Hamiltonian. In all these diagrams we neglect the contributions of the Pomeron loops or, in other words, we consider 
 the diagrams that contribute to the scattering amplitude in the kinematic region   $
G^2_{3\pom}\,e^{2\,\Delta_{\mbox{\tiny BFKL}} Y} \,\,\ll\,\,1$, while $ A_i\,G^2_{3\pom}\,e^{ \Delta_{\mbox{\tiny BFKL}}Y} \sim\,1$. $\Delta_{\mbox{\tiny BFKL}}$ is the intercept of the BFKL Pomeron. }
\label{netset}
\end{figure}

\begin{boldmath}
\section{BFKL Pomeron calculus in leading order of $N_c$ -  a recap}
\end{boldmath}
The BFKL Pomeron interaction in leading order of $N_c$ can be written as the following path integral \cite{BRAUN,KLL}:

\begin{equation} \label{BFKLFI}
Z[\Phi, \Phi^+]\,\,=\,\,\int \,\,D \Phi\,D\Phi^+\,e^S \,\,\,\hspace{0.5cm}\mbox{with}\hspace{0.5cm}\,S \,=\,S_0
\,+\,S_I\,+\,S_E
\end{equation}
where $S_0$ describes free Pomerons, $S_I$ corresponds to their mutual interaction
while $S_E$ relates to the interaction with the external sources (target and
projectile). $S_0 +  S_I$ takes the following form:
\beq \label{S0I} 
S_0 + S_I\,=\,\intl^Y_0\!\! d\eta \left[\frac{N_c^2}{4\pi^4\bar\alpha_s^2}[\nabla^2_x\nabla^2_y\bar\Phi(x,y,\eta)]\frac{\partial }{\partial\eta}\Phi(x,y, \eta)-H_B(\Phi ,\bar\Phi)\right]
\eeq
with $H_B\Lb \Phi,\bar{\Phi}\Rb$:
\bea \label{HB} 
H_B\Lb \Phi,\bar{\Phi}\Rb&=&\frac{N_c^2}{2\pi\bar\alpha_s}\intl\underbrace{\bar \Phi(x,y,\eta)\nabla^2_x\nabla^2_y\Bigg[K(x,y|z)\Lb\Phi(x,z, \eta)+\Phi(z,y, \eta)- \Phi(x,y, \eta)\Rb\Bigg]}_{H_{BFKL}}\nn\\
&-&\underbrace{\bar \Phi(x,y,\eta)\nabla^2_x\nabla^2_y\Bigg[K(x,y|z)\Phi(x,z, \eta)\Phi(z,y, \eta)\Bigg]-\Phi(x,y,\eta)\nabla^2_x\nabla^2_y\Bigg[
K(x,y|z)\bar\Phi(x,z, \eta)\bar\Phi(z,y,\eta)\Bigg]}_{H_I}
\eea
where   $\vec{x}$ and $\vec{y}$  are the positions of the quark and antiquark\footnote{The alternative notations: the dipole size $\vec{r} = \vec{x} - \vec{y}$ and the impact parameter $\vec{b} = \h \Lb \vec{x} + \vec{y}\Rb$.} of the dipole at rapidity $\eta$ and 
\beq \label{K}
 K\Lb \vec{x},\vec{y} | \vec{z}\Rb\,\,=\,\,\frac{
\Lb \vec{x}- \vec{y}\Rb^2}{\Lb \vec{x}- \vec{z}\Rb^2\,\Lb \vec{z}- \vec{y}\Rb^2}
\eeq
The equations of motion take the form:
\beq \label{EQA1}
\frac{\delta S}{\delta \Phi\Lb x,y\Rb}\,\,=\,\,0;\,\,\,\,\,\,\,\,\frac{\delta S}{\delta \bar \Phi\Lb x,y\Rb}\,\,=\,\,0;
\eeq
 The interpretation of $\Phi$ and $\bar \Phi$ is that of the scattering amplitude of an external dipole on the target and the projectile, respectively. For the dense-dense dipole system scattering (nucleus-nucleus interaction) \eq{EQA1}
  sums the net diagrams of \fig{netset}-b (see Refs.  \cite{BRAUN,KLL}). It should be noted that for the case of dipole-nucleus scattering the sum of net diagrams degenerates to the sum of 'fan' diagrams (see \fig{netset}-a) which can be sum by the  BK equation. This equation   is  the equation of motion for the  BK Hamiltonian. One can see that the equations of motion lead to the scattering amplitudes in the mean field approximation.

 To solve \eq{EQA1} we need to add the boundary conditions which are
\beq \label{PHI0}
\Phi_{\eta=0}(x,y)=\phi(x,y)\equiv\intl_{u,v}\gamma(x,y;u,v)J_T(u,v); \ \ \ \ \ \ \bar\Phi_{\eta=Y}(x,y)=\bar\phi(x,y)\equiv\intl_{u,v}\gamma(x,y;u,v)J_P(u,v)
\eeq 
with  the source terms $J_P(x,y)\delta(\eta-Y)$ for $\bar \Phi$ 
and  $J_T(x,y)\delta(\eta)$ for $\Phi$.

The classical equations of motion (see \eq{EQA1}) are given by
\begin{subequations}
\bea
 \,\,\,\,\,& & \frac{\partial \Phi(x,y;\eta)}{\partial\,\eta}\,\,= \label{BAEQ1}\\
 &=&\,\,\frac{\bas}{2
\pi}\,\int\,d^2\,z\,K\left(x,y|z
\right)\,\Big\{ \Phi(x,z;\eta) \,+\,\Phi(z,y;\eta)\,-\,\Phi(x,y;\eta)\,- \Phi(z,y;\eta)\,\Phi(x,z;\eta) \Big\} \nonumber
\\
&-& \,\frac{\bas}{ 2\pi}\,\int_{z,x',y'}L^{-1}_{xy;x'y'} \,\,K \left(x',y'|z
\right)\,\,\left[\left\{L_{zy'} \,\Phi\left(z,y',\eta\right)\right\}\,\bar \Phi(x',z; \eta)+\left\{L_{zx'} \,\Phi\left(z,x',\eta\right)\right\}\,\bar \Phi(y',z; \eta)\right]\,\nonumber\\
\nn\\
\nn\\
 \,\,\,\,\,& & - \frac{\partial \bar \Phi(x,y;\eta)}{\partial\,\eta}\,\,= \label{BAEQ2}\\
 &=&\,\,\frac{\bas}{2
\pi}\,\int\,d^2\,z\,K\left(x,y|z
\right)\,\Big\{ \bar \Phi(x,z;\eta) \,+\,\bar \Phi(z,y;\eta)\,-\,\bar \Phi(x,y;\eta)\,- \bar \Phi(z,y;\eta)\,\bar \Phi(x,z;\eta) \Big\} \nonumber
\\
&-& \,\frac{\bas}{ 2\pi}\,\int_{z,x',y'}L^{-1}_{xy;x'y'} \,\,K \left(x',y'|z
\right)\,\,\left[\left\{L_{zy'} \,\bar \Phi\left(z,y',\eta\right)\right\}\, \Phi(x',z; \eta)+\left\{L_{zx'} \,\bar \Phi\left(z,x',\eta\right)\right\}\, \Phi(y',z; \eta)\right]\,\nonumber
\eea
\end{subequations}
The operator $L_{xy} \,=\,(x - y)^4\nabla^2_x \,\nabla^2_y$. $L^{-1}_{xy;x'y'}  =\frac{1}{(2 \pi)^4}
{(x' - y')^4} G_0\left(x,y;Y'|x',y';Y'\right)$, where $G_0$
is equal to the initial Green function  of the BFKL Pomeron that corresponds to the exchange of two gluons. It has the following form

\begin{equation} \label{G0}
G_0(x_1,x_2;Y| x'_1,x'_2;Y)\,\,=\,\, \pi^2\,\ln\Lb
\frac{x^2_{1,1'}\,x^2_{2,2'}}{x^2_{1,2'}\,x^2_{1',2}}\Rb  \,\ln
\Lb\frac{x^2_{1,1'}\,x^2_{2,2'}}{x^2_{1,2}\,x^2_{1',2'}}\Rb
\end{equation}
where $\vec{x}_1 = \vec{x}, \vec{x}_2 = \vec{y}$. 
 This form of $G_0$ has been discussed in Ref.\cite{LIP}. $G_0$ satisfies
\beq\label{G01}
\nabla^2_x\nabla^2_yG_0(x,y,Y|x',y',Y)= (2\,\pi)^4\,\,[\delta^2(\vec{x}-\vec{x}')\delta^2(\vec{y} - \vec{y}')+
\delta^2(\vec{x}-\vec{y}')\delta^2(\vec{y} - \vec{x}')]\eeq

~

~

\begin{boldmath}
\section{Solutions in the vicinities of the fixed points}
\subsection{Perturbative QCD region}
\end{boldmath}
For small $\Phi$ and $\bar \Phi$ \eq{BAEQ1} and \eq{BAEQ2} can be reduced to the linear BFKL equations:
\beq \label{SOL1}
\frac{\partial \Phi(x,y;\eta)}{\partial\,\eta}\,\,= \,\,\frac{\bas}{2
\pi}\,\int\,d^2\,z\,K\left(x,y|z
\right)\,\Big\{ \Phi(x,z;\eta) \,+\,\Phi(z,y;\eta)\,-\,\Phi(x,y;\eta) \Big\}
\eeq
The eigenfunctions of these equations have the following form\cite{LIP}:
\begin{subequations}
\bea
\phi_\gamma\Lb \vec{r}, \vec{r}_P, \vec{b}\Rb &=& \Lb \frac{r^2\,r^2_P}{\Lb \vec{b} - \h\Lb \vec{r}-\vec{r}_P\Rb\Rb^2 
\,\Lb \vec{b} + \h\Lb \vec{r}-\vec{r}_P\Rb\Rb^2 }\Rb^\gamma\,\,e^{\gamma\,\xi_{r,r_P}}; \label{EF}\\
\mbox{with}~~ \xi &=& \ln\Lb \frac{r^2\,r^2_P}{\Lb \vec{b} - \h\Lb \vec{r}-\vec{r}_P\Rb\Rb^2 
\,\Lb \vec{b} + \h\Lb \vec{r}-\vec{r}_P\Rb\Rb^2 }\Rb \label{XI};
\eea
\end{subequations}
Actually, in Ref.\cite{LIP} it is proven that this eigenfunction is correct for any kernel that satisfies the conformal symmetry. The eigenvalue for the BFKL equation has the form:
\beq \label{EV}
\omega\Lb \bas, \gamma\Rb \,\,=\,\,\bas \chi\Lb \gamma\Rb = \bas\Lb 2 \psi(1) - \psi\Lb \gamma\Rb  -  \psi\Lb 1 - \gamma\Rb\Rb
\eeq
where $\psi(z)$ denotes the Euler psi-function: $\psi(z) = d \ln \Gamma\Lb z\Rb/d z$. In \eq{BAEQ1}
 and      \eq{BAEQ2} we denote  by $\vec{x} $ and $\vec{y}$ the coordinates of the quark and antiquark in the colourless dipole, respectively. In \eq{EF} and    \eq{XI} we introduce: (i) the size of the dipole: $\vec{r} = \vec{x} - \vec{y}$, (ii) the size of the dipole in the projectile (or target) $\vec{r}_P = \vec{x}_P - \vec{y}_P$ and (iii) the impact parameter $\vec{b} = \h\Lb \vec{x} +\vec{y} - \vec{x}_P - \vec{y}_P\Rb$. 
 
 $\Phi\Lb x,y,\eta\Rb$ increases with $\eta$ and when it becomes large enough,  the shadowing corrections become essential. The solution to the BFKL equation has the following form \cite{MUT}:
 \beq \label{MTEQ}
 \Phi\Lb x,y,\eta\Rb\,\,=\,\,\phi_0 e^{\bg\,z} ~~\mbox{with}~~z\,\, = \,\,\bas \chi\Lb \gamma_{cr}\Rb \,\eta + \xi_{r,r_P}
 \eeq 
 in the vicinity of the saturation scale ($z \to 0$). $\bg = 1 - \gamma_{cr}$  is the solution to the following equation:
  \beq \label{GACR}
\kappa \,\,\equiv\,\, \frac{\chi\Lb \gamma_{cr}\Rb}{1 - \gamma_{cr}}\,\,=\,\, - \frac{d \chi\Lb \gamma_{cr}\Rb}{d \gamma_{cr}}~
\eeq 
 
  For  $\bar \Phi\Lb x,y,\eta\Rb$  the BFKL equation leads to the following solution at $\bar{z} \to 0$:
 \beq \label{MTEQ1}
 \bar \Phi\Lb x,y,\eta\Rb\,\,=\,\,\bar{\phi}_0 e^{\bg \,\bar{z}} ~~\mbox{with}~~\bar{z}\,\, = \,\,\bas \chi\Lb \gamma_{cr}\Rb \,\Lb Y\,-\,\eta\Rb + \xi_{r,r_T}
 \eeq  
 Both constants  $\phi_0 $ and $\bar{\phi}_0$ are determined by the initial conditions to the BFKL equation at $\eta=0$ and $\eta =Y$ for $\Phi$ and $\bar \Phi$, respectively. 
 
\begin{figure}[ht]
    \centering
  \leavevmode
     \includegraphics[width=7cm]{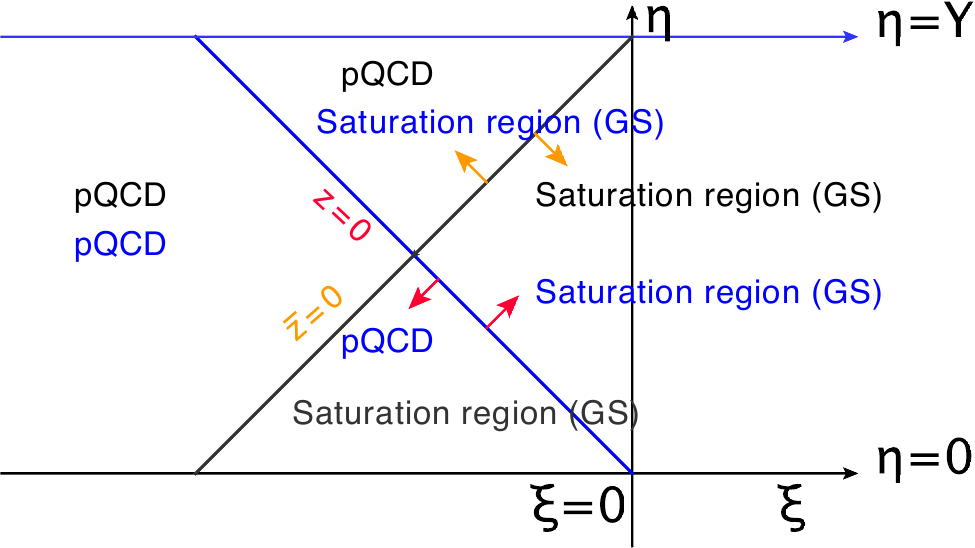}  
      \caption{Kinematic regions for $\Phi$ (blue colour)  and $\bar \Phi$ (black colour).  }
\label{brhgen}
   \end{figure}

 We see that at small $z$ and $\bar{z}$ both $\Phi$ and $\bar \Phi$ show the geometric scaling behaviour, each  being a function of one variable $z$ or $\bar{z}$. We assume that the general solutions in the saturation region for $z\,>\,0$ and $\bar z\,>\,0$ take the forms   $\Phi\Lb z\Rb$ and $\bar{\Phi}\Lb \bar z\Rb$. In \fig{brhgen} we show the different regions for $\Phi$ and $\bar{\Phi}$ using GS as a notation for the geometric scaling behaviour\cite{GS}.

Generally speaking the system of equations (see \eq{BAEQ1} and \eq{BAEQ2}) have four fixed point  $\Lb \Phi,\bar\Phi\Rb$: (0,0),(1,0),(0,1) and (1,1). We have discussed
 the first fixed point in this section and  have seen that this is an  unstable fixed point since the solutions of the BFKL equation lead to an increase of both $\Phi$ and $\bar \Phi$. 

~

~

\begin{boldmath}
\subsection{  Both $\Phi$ and $\bar \Phi$  are far away from saturation.}
\end{boldmath}

\begin{figure}
\begin{center}
\includegraphics[width=8cm]{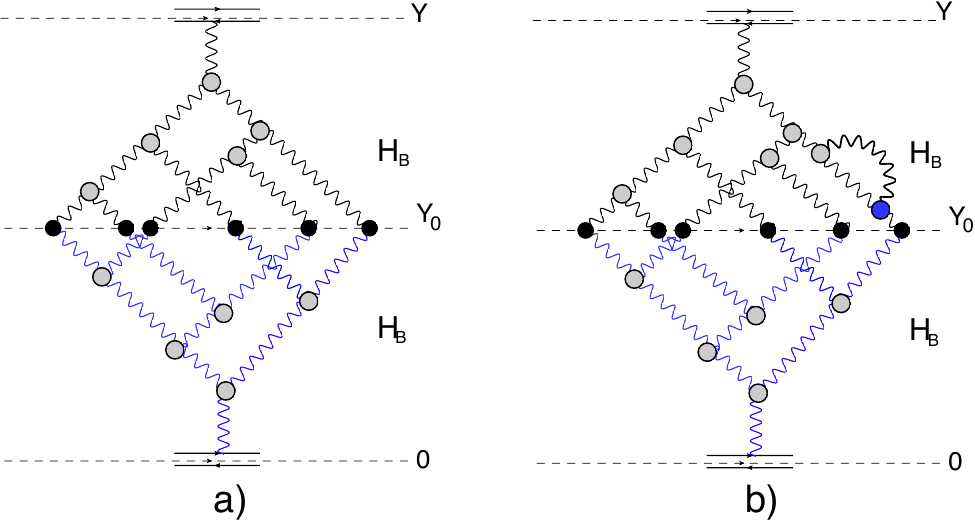}
\end{center}
\caption{\label{cascade} Summing large Pomeron loops.
The wavy lines denote the BFKL Pomerons. The grey circles denote  the triple Pomeron vertices, while the black circle denote the amplitude of two dipoles interaction in the Born approximation of perturbative QCD.}
\label{lpoml}
\end{figure}

 As  we have seen in the previous subsection, both $\Phi$ and $\bar \Phi$  increase with the growth of energy.  In the BFKL Pomeron calculus this increase is shown as the `fan' Pomeron diagrams in \fig{lpoml}-a. These diagrams describe  only the increase of the dipole densities since they depend on the triple Pomeron vertex $ \pom \to 2\,\pom$.  The inverse process of reducing  the dipole densities is related to the Pomeron loops diagrams. These diagrams depend on the vertices $\pom + \pom \to \pom$ and the simplest one is shown in \fig{lpoml}-b.  Certainly, such diagrams can be essential only for a dense system of dipoles. A simple estimate    shows that the contributions of the loop diagrams are suppressed by factor  
 \beq \label{SPSPB1}
 \bas^2 e^{\Delta_{\mbox{\tiny BFKL}} \,\eta}\,\, <\,\,1
 \eeq
 where  $\Delta_{\mbox{\tiny BFKL}}$ is the intercept of the BFKL Pomeron  (see Ref.\cite{AKLL} for details).

 Therefore, for $\eta \,<\,\eta_{max} = \frac{1}{\Delta_{\mbox{\tiny BFKL}}} \ln \Lb \frac{1}{\bas^2}\Rb$ we can safely neglect the contribution of the Pomeron loops. In other words,  it means that for 
   \beq \label{SPSPB2}
Y \,\, \leq\,\,2 \,\eta_{max}
 \eeq 
 we can safely use the diagrams of \fig{lpoml}-a  in  calculating the dipoles densities. Therefore, for a  large range in rapidity the sum of large Pomeron loops gives the scattering amplitude.  We will discuss this approach in  section IV in more detail.

Intuitively, we believe that the behaviour of $\Phi$ and $\bar \Phi$ in the vicinity of the fixed point (1,1) corresponds to the scattering  amplitudes at high energies.

~

~

\begin{boldmath}
\subsection{Saturation regions for both $\Phi$ and $\bar \Phi$}

\subsubsection{Equation of motion}
\end{boldmath}
In the saturation regions for both $\Phi$ and $\bar \Phi$ (see \fig{brhgen})  we expect that the behaviour of these functions in the vicinity of the fixed point (1,1)  will give the high energy asymptotic of the scattering amplitudes.
Following Ref.\cite{LETU} we introduce new functions $\Delta$ and $\bar \Delta$:  $ \Phi\Lb x,y,\eta\Rb\,=1 - \Delta\Lb x,y,\eta\Rb$ and  $\bar  \Phi\Lb x,y,\eta\Rb\,=1 - \bar\Delta\Lb x,y,\eta\Rb$. We assume that both $\Delta $ and $\bar \Delta$ are small ($\Delta \ll 1$ and $ \bar\Delta \ll 1$). Neglecting all contributions that are proportional to $\Delta^2$, $\Delta \bar \Delta$ and $\bar{\Delta}^2$,  we obtain two independent linear equations \cite{KLL}  for $\Delta $ and $\bar \Delta$:
\bea
&&\hspace{-0.5cm} \frac{\partial \Delta(x,y;\eta)}{\partial\,\eta}= -\frac{\bas}{2
\pi}\,\Big\{\intl_z\!\!K\left(x,y|z\right)\,\,\Delta(x,y;\eta)\,+\intl_{zx'y'}\!\!\!\!\!L^{-1}_{xy;x'y'} \,\,K \left(x',y'|z
\right)\,\,\left\{L_{zy'} \,\Delta\left(z,y',\eta\right)+L_{zx'} \,\Delta\left(z,x',\eta\right)\right\} \Big\} \label{1AEQ1}\\
&&\hspace{-0.5cm}-\frac{\partial \bar\Delta(x,y;\eta)}{\partial\,\eta}=\frac{\bas}{2
\pi}\,\Big\{\intl_z\!\!K\left(x,y|z\right)
\,\,\bar\Delta(x,y;\eta)\,+\intl_{zx'y'}\!\!\!\!\!L^{-1}_{xy;x'y'}\,K \left(x',y'|z
\right)\,\,\left\{L_{zy'} \,\bar\Delta\left(z,y',\eta\right)+L_{zx'} \,\bar\Delta\left(z,x',\eta\right)\right\}\Big\}\,\label{2AEQ2}
\eea
These equation can be simplified by multiplying  them by the operator $L$ and rewriting then for  functions $n(x, y, \eta)=L_{xy}\Delta(x,y,\eta)$ and $\bar n(x,y)=L_{xy}\bar \Delta(x,y,\eta)$ .
\bea
 \frac{\partial n(x,y;\eta)}{\partial\,\eta}\,\,
 &=&\,\,-\frac{\bas}{2
\pi}\,\Big\{\int_z\,\,K\left(x,y|z\right)\,\,n(x,y;\eta)\,+\int_{z}\, \,\,K \left(x,y|z
\right)\,\,\left\{n\left(z,y,\eta\right)+n\left(z,x,\eta\right)\right\} \Big\} ; \label{1AEQ12}\\
 - \frac{\partial \bar n(x,y;\eta)}{\partial\,\eta}\,\,
&=&\,\,-\frac{\bas}{2
\pi}\,\Big\{\int_z\,K\left(x,y|z\right)
\,\,\bar n(x,y;\eta)\,+\int_z\,\,K \left(x,y|z
\right)\,\,\left\{\bar n\left(z,y,\eta\right)+ \,\bar n\left(z,x,\eta\right)\right\}\Big\}; \label{1AEQ22}\eea
Assuming the geometric scaling behaviour of $n$ and $\bar n$ in this region one can see that these equations take the form:
\bea
\kappa  \frac{d n(z)}{d\,z}\,\,
 &=&\,\,-2\,z\,\,n(z)\,-\int_{z}\, \,\,K \left(x,y|z
\right)\,\,\left\{n\left(z,y,\eta\right)+n\left(z,x,\eta\right)  - n\left(x,y,\eta\right )\right\} ; \label{NF}\\
 \kappa  \frac{d\bar n(\bar z)}{d\,\bar z}\,\,
&=&\,\,-2\,\bar z\,\,\bar n(\bar z)\,\,- \,\int_{z}\, \,\,K \left(x,y|z
\right)\,\,\left\{\bar n\left(z,y,\eta\right)+ \bar n\left(z,x,\eta\right)  - \bar n\left(x,y,\eta\right )\right\} ; \label{NBF}
\eea
$z$ and $\bar z$ are given by \eq{MTEQ} and \eq{MTEQ1}, respectively \footnote{We hope that the notation $z$ for  the energy variable and for the  coordinate will not be misleading.}. $\kappa $ and $\gamma_{cr}$ are determined by \eq{GACR} \cite{GLR,MUT}.

The general solution to \eq{NF} and \eq{NBF} can be obtained by using the Mellin transforms:
\beq \label{SOL2}
n(z)\,\,=\,\,\intl^{\epsilon + i \infty}_{ \epsilon - i\infty} \frac{d \lambda}{2\,\pi\,i} e^{\lambda\,z} \Phi\Lb \lambda\Rb;~~~
\bar n(z)\,\,=\,\,\intl^{\epsilon + i \infty}_{\epsilon - i \infty} \frac{d \lambda}{2\,\pi\,i}e^{\lambda\,z} \bar\Phi\Lb \lambda\Rb;
\eeq
For $\Phi\Lb \lambda\Rb$ we have the following equation:
\beq \label{SOL3}
\lambda \,\kappa\,\Phi\Lb \lambda\Rb\,\,=\,\, 2\,\frac{d}{d \lambda}\Phi\Lb \lambda\Rb\,\,-\,\,\chi\Lb \lambda\Rb \Phi\Lb \lambda\Rb
\eeq
with the solution:
\beq \label{SOL4}
\Phi\Lb \lambda\Rb= C\,\exp\Lb \frac{1}{4} \kappa\,\lambda^2 \, +\,\h\intl^\lambda_0 d \lambda' \chi\Lb \lambda'\Rb\Rb
\eeq
In \eq{SOL4} the constant $C$ has to be determined from the initial conditions.

Plugging this solution into \eq{SOL2} we get 
\beq \label{SOL5}
n(z)=\intl^{\epsilon + i \infty}_{ \epsilon - i\infty} \frac{d \lambda}{2\,\pi\,i} \exp\Lb\lambda\,z +\frac{1}{4} \kappa\,\lambda^2 \, +\,\h\intl^\lambda_0 d \lambda' \chi\Lb \lambda'\Rb\Rb\,=\,\intl^{\epsilon + i \infty}_{ \epsilon - i\infty} \frac{d \lambda}{2\,\pi\,i} \exp\Lb\lambda\,z +\frac{1}{4} \kappa\,\lambda^2 \,+\h \psi(1)\lambda \Rb
\sqrt{\frac{\Gamma\Lb 1 - \lambda\Rb}{\Gamma\Lb \lambda\Rb}}
 \eeq 
For large $z$ we can calculate this integrals using the method of steepest descent with the following solution for the saddle point $\lambda_{SP}$:
\beq \label{SOL6}
\h\, \kappa\, \lambda_{SP}\,\, +\,\, z \,\,+ \,\,\h \chi\Lb \lambda_{SP} \Rb = 0
\eeq
The main contribution stems from large $\lambda$ where \eq{SOL6} takes the following form:
\beq \label{SOL7}
\h\, \kappa\, \lambda_{SP}\,\, +\,\, z \,\, - \ln  \lambda_{SP} = 0
\eeq
with the solution:
\beq \label{SOL8}
\lambda_{SP} = - \frac{2 \,z}{\kappa} + \ln\Lb  \frac{2 \,z}{\kappa}\Rb
\eeq
Using this $\lambda _{SP}$ we can  obtain the following solution for the equation of motion:
\bea\label{SOL9}
&&n^{EOM}(z)=\\
&&C\exp\Lb - \frac{ \Lb z - \h \kappa \ln\Lb  \frac{2 \,z}{\kappa}\Rb\Rb^2}{\kappa }  + \h\intl^{\lambda_{SP}}_0 \!\!\!d \lambda' \chi\Lb \lambda'\Rb\Rb = C \exp\Lb - \frac{ \Lb z - \h \kappa \ln\Lb  \frac{2 \,z}{\kappa}\Rb\Rb^2}{\kappa } +       \h \psi(1)\lambda_{SP} \Rb
\sqrt{\frac{\Gamma\Lb 1 - \lambda_{SP}\Rb}{\Gamma\Lb \lambda_{SP}\Rb}}\nn
\eea
One can see that $n^{EOM}(z)\,\,\propto \,\,\exp\Lb - \frac{z^2}{\kappa }\Rb\,\,\ll\,\,1$ at $z>1$. The same behaviour holds  for $\bar n$. Therefore, the fixed point (1,1) is stable. 

~

\begin{boldmath}
\subsubsection{Scattering matrix}
\end{boldmath}

The scattering matrix is given by the following equation\cite{BRAUN,KLL}:\beq\label{action}
{\cal S}\simeq \int d \bar\Phi \,d\Phi \,W_P[\Phi]\,W_T[\bar\Phi]\!\!\!\!\!\!\!\!\!\!\!\!\!\!
\intl_{\Phi(Y)=\Phi;\ \ \bar \Phi(0)=\bar\Phi}\!\!\!\!\!\!\!\!\!\!\!\!\!\!\!\!D\Phi(\eta)\,D\bar \Phi(\eta) e^{S_0+S_I}\Bigg{/} 
\intl D\Phi(\eta)\,D\bar \Phi(\eta) e^{S_0+S_I}
\eeq
with
\beq\label{action0}\
S_0 + S_I \,=\,\int^Y_0 d\eta \left[\frac{N_c^2}{4\pi^4\bar\alpha_s^2}[\nabla^2_x\nabla^2_y\bar\Phi(x,y; \eta)]\frac{\partial }{\partial\eta}\Phi(x,y; \eta)-H(\Phi ,\bar\Phi)\right]
\eeq
and
\beq \label{action1}
W_P=\exp[-\int J_P(x,y)\Phi(x,y)]; \ \ \ \ \ W_T=\exp[-\int J_T(x,y)\bar\Phi(x,y)]\eeq
where the initial conditions for $\Phi $ and $\bar \Phi$  have been discussed in \eq{PHI0}. For the case of dipole-dipole scattering \eq{action} reduces to the following equation (see also Eq.2.5 and Eq.2.6 of Ref.\cite{KLL}):
\beq\label{action2}
{\cal S}\simeq \Lb 1 -  \Phi\Lb \vec{x} , \vec{y}; \eta = Y\Rb\Rb \,\Lb 1 -  \bar\Phi\Lb \vec{x} , \vec{y}; \eta=0\Rb\Rb
\!\!\!\!\!\!\!\!\!\!\!\!\!\!
\intl_{\Phi(Y)=\Phi;\ \ \bar \Phi(0)=\bar\Phi}\!\!\!\!\!\!\!\!\!\!\!\!\!\!\!\!D\Phi(\eta)\,D\Phi(\eta) e^{S_0+S_I}\Bigg{/} 
\intl D\Phi(\eta)\,D\bar \Phi(\eta) e^{S_0+S_I}
\eeq
The equations of motion, which  we have discussed above, give the saddle point approximation for $\Phi^{EOM}(x,y; \eta)$ and $\bar \Phi^{EOM}\Lb x, y; Y -  \eta\Rb$ in \eq{action2}. For these solutions the contribution of the action is equal to  
\beq \label{action21}
S_0 \Lb \Phi^{EOM},\bar \Phi^{EOM}\Rb+S_I \Lb \Phi^{EOM},\bar \Phi^{EOM}\Rb= 2\intl^Y_0 d \eta \Delta(x,y,\eta)\nabla^2_x\nabla^2_y\Bigg[
K(x,y|z)\bar\Delta(x,z, \eta)\Bigg]
\eeq
resulting in a path    integral contribution that turns out to be small at
large $Y$.
\beq \label{action22}
 2\intl^Y_0 d \eta\Delta(x,y,\eta)\nabla^2_x\nabla^2_y\Bigg[
K(x,y|z)\bar\Delta(x,z,Y -  \eta)\Bigg]\,\,\xrightarrow{z\gg 1}\,\, \exp\Lb - \frac{z^2}{\kappa} \Rb\,\,\ll\,1
\eeq
Therefore, the contribution of the integral  $\intl^Y_0 d \eta(S_0+S_I)$ is negligible and 
 the scattering matrix in the vicinity of the fixed point (1,1)  is given by 
\bea\label{action3}
{\cal S}&\simeq& \Lb 1 -  \Phi\Lb \vec{x} , \vec{y}; \eta = Y\Rb\Rb \,\Lb 1 -  \bar\Phi\Lb \vec{x} , \vec{y}; \eta=0\Rb\Rb\,\,=\,\, \Delta\Lb\vec{x} , \vec{y}; \eta = Y\Rb\,   \bar\Delta\Lb \vec{x} , \vec{y}; \eta=0\Rb \nn\\
&  \sim&\,\,\Psi\Lb z \Rb\exp\Lb  - 2\frac{ \Lb z - \h \kappa \ln\Lb  \frac{2 \,z}{\kappa}\Rb\Rb^2}{\kappa }\Rb
\eea
where $\Psi$ is a smooth function of $z$.

\eq{action3} is in striking contrast to  the estimates for S-matrix in `rare' fluctuation approach\cite{IAMU,IAMU1} as well as with the results of summing  the large BFKL Pomeron loops\cite{LELT,LE1,LEDIDI,LEDIA,LEAA}. The question of
in which kinematic region we can trust our procedure of summing the large
Pomeron loops will be addressed in the next section.
 Here we wish to stress that \eq{action3} is derived on the same footing as the asymptotic behaviour\cite{LETU}  of the scattering amplitude for dipole-nucleus interaction which satisfies the BK equation. Both the Braun Hamiltonian and the BK one have the same problems with s-channel unitarity\cite{KLL,KLLL1,KLLL}, but we  believe that these problems  will not affect the asymptotic behaviour of the scattering amplitude at high energies.

However, before discussing these issues we wish to note
that  the smooth function $\Psi$ in \eq{action3}  is rather large. The question arises whether we can trust  our calculation of this function. To investigate this problem  we  are going to clarify this  behaviour of $n\Lb z\Rb$ using
the simplified BFKL kernel, which takes into account only the leading twist contributions. 
 This kernel describes the high energy asymptotic solution of the nonlinear BK equation and leads to the geometric scaling behaviour: 
    \bea \label{SIMKER}
\chi\Lb \gamma\Rb\,\,=\,\, \left\{\begin{array}{l}\,\,\,\frac{1} {1\,-\,\gamma}\,\,\,\,\,\,\,\,\,\,\mbox{for}\,\,z = \xi_{r,R} + \kappa\,\bas\,Y \,>\,0,\,\,\,\,\,\,\mbox{summing} \Lb z\Rb^n \mbox{terms};\\ \\
\,\,\,\frac{1}{\gamma}\,\,\,\,\,~~~~~~\mbox{for}\,\,\,z \,<\,0,\,\,\,\,\,~~~~~~\,\,\,\,\,~~~~~~~~~~~~~\mbox{summing}
\Lb \xi\Rb^n \mbox{ terms};\\  \end{array}
\right.
\eea
 where  $\kappa = 4 $ for this kernel.

 Since this kernel sums log contributions, it corresponds to  the  leading twist   term of the full BFKL kernel. It has a very simple form in the coordinate representation \cite{LETU}:
 
\beq \label{K2}
\intl \, \displaystyle{ K\Lb \vec{r}',\vec{r} - \vec{r'}|\vec{r}\Rb}\,d^2 r' \,\rightarrow
\,\frac{\bas}{2}\!\!\!\!\! \!\!\!\!\! \intl^{r^2}_{1/Q^2_s(Y,b)}\!\!\!\!\!\!  \frac{ d r'^2}{r'^2}\,\,+\,\,
\frac{\bas}{2}\!\!\!\!\! \!\!\!\!\! \intl^{r^2}_{1/Q^2_s(Y, b)}\!\!\!\!\!\!\! \frac{ d |\vec{r} - \vec{r}'|}{|\vec{r}  - \vec{r}'|^2}\,\,
 =\,\,\frac{\bas}{2}\, \intl^{\xi}_{\xi_s} d \xi_{r'} \,\,+\,\,\frac{\bas}{2}\, \intl^{\xi}_{\xi_s} d \xi_{\vec{r} - \vec{r}'}\eeq
 where $\xi_{r'} \,=\,\ln \Lb \frac{r'^2\,r^2_1}{b^4}\Rb$ for $b \,>\,r,r_1$
 and $\xi_s\,=\,\kappa\, \bas\,Y$ for the scattering of the  dipole $r'$ with the dipole $r_1$.
 Note, that  the  logarithms
   originate from the decay of a large size dipole into one small-
 size dipole  and one large size dipole\cite{LETU}.  However, the size of the
 small dipole is still larger than $1/Q^2_s  $. 
 
 \eq{NF} reduces to the following equation:
 \beq \label{SOL10}
 \kappa  \frac{d n(z)}{d\,z}\,\,
=\,\,-2\,z\,\,n(z)\,-\int^z_0 d z' n(z')
\eeq
Following the procedure discussed in \eq{SOL5}-\eq{SOL9} we obtain at large $z$ that
 \beq \label{SOL11}
 n(z)\,\,=\,\, C \sqrt{z}\,\,\exp\Lb - \frac{z^2}{\kappa}\Rb
 \eeq
Now let us estimate $\Delta^2$ corrections by considering the first line of \eq{BAEQ1}. For the leading twist kernel it takes the form:
 \beq \label{SOL12}
- \,z\,\Delta(z) + \Delta(z) \intl^z_0 d z' \,\Delta(z') = - \,(z\,-\,z_0)\,\underbrace{\Delta(z)}_{\exp\Lb - \frac{z^2}{4 \,\kappa}\Rb}
 - \underbrace{\Delta(z) \intl^\infty_z d z' \,\Delta(z')}_{exp\Lb - \frac{z^2}{2 \,\kappa}\Rb} 
\eeq
with $z_0 = \intl^\infty_0 d z' \Delta(z')$.
Replacing in the second line of \eq{BAEQ1} $\intl^{\bar z}_0 d \bar z' \bar\Delta(\bar z') $    by   $ \intl^{\infty}_0 d \bar z' \bar\Delta(\bar z') \,-\,  \intl^{\infty}_{\bar z} d \bar z' \bar\Delta(\bar z')   $
we can see that the term $ \Delta \bar \Delta$ is small. Therefore, these estimates show that we can guarantee that any smooth function $\Psi  \,<\,\exp\Lb - \frac{z^2}{4 \,\kappa}\Rb$ at large $z$   can be found from the solution of \eq{1AEQ12} and \eq{1AEQ22}.
The only changes that we need to make in \eq{NF} and \eq{NBF}  are $ z \to z - z_0 = z - \h\Lb \int^{\infty}_0 d z' \Delta(z') + \int^{\infty}_0 \bar z' \bar \Delta(\bar z')\Rb$. 

Concluding this section, we state that (i) the fixed point (1,1) is stable and $n$ and $\bar n$ in the vicinity of this point can be found from \eq{SOL9} and  (ii) \eq{SOL9} can be trusted for $\Psi\, < \,\exp\Lb - \frac{z^2}{\kappa}\Rb$.

~

~

\begin{boldmath}
\subsection{ $\Phi$ and $\bar \Phi$ in the vicinity of the fixed point (1,0)}
\end{boldmath}
 In the vicinity of this point 
 $1-\Phi\equiv \Delta,\ \ \ \bar \Phi\equiv \bar\Delta$ with small $\Delta$ and $\bar \Delta$. The equations take the form
\begin{subequations}
\bea
 \frac{\partial \Delta(x,y;\eta)}{\partial\,\eta} &=&-\,\,\frac{\bas}{2
\pi}\,\int_z\,\, K\left(x,y|z
\right)\,\,\Delta(x,y;\eta); \label{2AEQ1}
\\
 - \frac{\partial \bar\Delta(x,y;\eta)}{\partial\,\eta}\,\,
 &=&\,\,\frac{\bas}{2
\pi}\,\int_z\,d^2 z\,K\left(x,y|z
\right)\,\Big\{ \bar\Delta(x,z;\eta) \,+\,\bar\Delta(z,y;\eta)\,-\,\bar\Delta(x,y;\eta)\, \Big\} \nonumber \\
&-& \,\frac{\bas}{ 2\pi}\int_z\,d^2z\,L^{-1}_{xy;x' y'}\,K \left(x',y'|z
\right)\,\,\left\{L_{z y'} \,\bar\Delta\left(z,y',\eta\right)+L_{zx'} \,\bar\Delta\left(z,x',\eta\right)\right\};\label{2AEQ2}
\eea
\end{subequations}

For $\bar n$ the second equation takes a very simple form:
\begin{equation}\label{NB2}
 \frac{\partial \bar n(x,y;\eta)}{\partial\,\Lb Y - \eta\Rb}\,\,= -\frac{\bas}{2
\pi}\,\int_z\,d^2 z\,K\left(x,y|z
\right)\,\,\bar n(x,y; Y - \eta)\,
\end{equation}

Both \eq{2AEQ1} and \eq{NB2} show that the functions $\Delta$ and $\bar n$ are small in the vicinity of this fixed point, making this fixed point  stable. The solution to these two equations have the form:
\begin{subequations}
\bea
\Delta(x,y;\eta) &=&\,\,{\rm Const} \exp\Lb - \frac{z^2}{2\,\kappa}\Rb; \label{2AEQ3}\\
 \bar  \Delta(x,y;\eta) \,\,&=&\,\,\Delta_0\exp\Lb - \bar \gamma\,z\Rb;
\label{2AEQ4}
\eea
\end{subequations}
One can see that  \eq{2AEQ3} is the solution to the BK equation of Ref.\cite{LETU} while the solution of \eq{2AEQ4} stems from \eq{MTEQ}.

 We can find the scattering matrix  from \eq{action2}. The Lagrangian for the solutions     to  the equation of motion is equal to zero and 
$S_0 \Lb \Phi^{EOM},\bar \Phi^{EOM}\Rb+S_I \Lb \Phi^{EOM},\bar \Phi^{EOM}\Rb$ vanishes at large $z$. Hence,
 at large $z$ we have
\beq \label{2AEQ5}
{\cal S} \simeq  \Lb \Delta\Lb \vec{x} , \vec{y}; \eta = Y\Rb\Rb \,\Lb 1 -  \bar \Delta \Lb \vec{x} , \vec{y}; \eta=0\Rb\Rb\,\,\xrightarrow\,{\cal S}_{BK}\,=\,\exp\Lb - \frac{z^2}{2\,\kappa}\Rb
\eeq
One can see that in the vicinity of the fixed point $(1,0)$ the scattering amplitude coincides with that  of the BK equation( see Ref.\cite{LETU})). 

 Therefore, depending on the initial condition, the scattering amplitude will be black for both projectile and target (fixed point (1,1)) or it will be black for the projectile but transparent for the target even at high energy. It goes against our intuition and , as it has been shown in Ref.\cite{KLL},  it contradicts QCD. In other words,  Braun Hamiltonian violates the s-channel unitarity. Unfortunately, in spite of intensive work it has not been suggested the Hamiltonian in QCD that satisfies both  $t$- and $s$- unitarity.

~

~

\begin{boldmath}
\section{ Summing large Pomeron loops }
\subsection{Generalities}
\end{boldmath}

 The way of summation stems from the  $t$-channel unitarity, which has been rewritten in a convenient form for the dipole approach to CGC in Refs.\cite{MUSA,IAMU,IAMU1,KOLEB,MUDI,LELU,KO1,LE11,AKLL}(see \fig{lpoml}-a).
 $N\Lb Y, r,R ;  \vec{b}\Rb $ is  the imaginary part of the scattering amplitude for the interaction  of the dipole with size $r$ and rapidity $Y$ at the impact parameter $b$ with the dipole of size $R$ which is at  rest ($Y=0$,$b=0$). For $N$
 the analytic expression takes the form
       \cite{LELU,KO1,LE1,AKLL}:  \bea \label{MPSI}
     && N\Lb Y, r,R ;  \vec{b}\Rb\,=\\
     &&\,\sum^\infty_{n=1}\,\Lb -1\Rb^{n+1}\,n!\int  \prod^n_i d^2 r_i\,d^2\,r'_i\,d^2 b'_i 
     \int \!\!d^2 \delta b_i\, \gamma^{BA}\Lb r_i,r'_i, \vec{b}_i -  \vec{b'_i}\equiv \delta \vec{b} _i\Rb 
    \,\,\rho_n\Lb Y - Y_0, \{ \vec{r}_i,\vec{b}_i\}\Rb\,\rho_n\Lb Y_0, \{ \vec{r}'_i,\vec{b}'_i\}\Rb \nn
      \eea
  $\gamma^{BA}$ is the scattering amplitude of two dipoles in the Born approximation of perturbative QCD.  The dipole densities $\rho_n\Lb Y , \{ \vec{r}_i,\vec{b}_i\}\Rb$ have been introduced in Ref.\cite{LELU}  as follows:
\beq \label{PD}
\rho_n(r_1, b_1\,
\ldots\,,r_n, b_n; Y\,-\,Y_0)\,=\,\frac{1}{n!}\,\prod^n_{i =1}
\,\frac{\delta}{\delta
u_i } \,Z\left(Y\,-\,Y_0;\,[u] \right)|_{u=1}
\eeq
  where  the generating functional $Z$ is
  \beq \label{Z}
Z\Lb Y, \vec{r},\vec{b}; [u_i]\Rb\,\,=\,\,\sum^{\infty}_{n=1}\int P_n\Lb Y,\vec{r},\vec{b};\{\vec{r}_i\,\vec{b}_i\}\Rb \prod^{n}_{i=1} u\Lb \vec{r}_i\,\vec{b}_i\Rb\,d^2 r_i\,d^2 b_i
\eeq
 where $u\Lb \vec{r}_i\,\vec{b}_i\Rb \equiv\,u_i$ is an arbitrary function and $P_n$ is the probability of having  $n$ dipoles with the  given kinematics.
 The initial and  boundary conditions for the BFKL cascade,  which stem from one dipole, have
the following form for the functional $Z$:
 
 \begin{subequations}
\bea
Z\Lb Y=0, \vec{r},\vec{b}; [u_i]\Rb &\,\,=\,\,&u\Lb \vec{r},\vec{b}\Rb;~~~~~~~~Z\Lb Y, r,[u_i=1]\Rb = 1;\label{ZIC}\\
\rho_1\Lb Y=0,   r,b, r_1,b_1\Rb\,\,&=&\,\,\delta^{(2)}\Lb \vec{r} - \vec{r}_1\Rb \delta^{(2)}\Lb \vec{b} - \vec{b}_1\Rb ;~~~~~\rho_n\Lb 
 Y=0, \vec{r},\vec{b}; [r_i, b_i]\Rb \,=\,0 ~ \mbox{at}~\,n\geq 2;\label{ZSR}
\eea
\end{subequations}
  In \eq{MPSI} $\vec{b}_i\,\,=\,\,\vec{b} \,-\,\vec{b'}_i$.

 As one can see from \fig{lpoml}-a the densities $\rho_n$ have to be found using the Braun  Hamiltonian. They have been discussed in Ref.\cite{LELT} in the saturation regions for the scattering amplitudes. The scattering amplitudes that have been found using \eq{MPSI}\cite{LELT,LE1,LEDIDI,LEDIA,LEAA} look reasonable,
but there are  two 
features which are not clear: whether this set of  diagrams satisfies the s-channel unitarity and whether it can describe the behaviour of the scattering amplitude at high energies. We obtain the  answer to the second question in the previous section and it is  negative. We can find the kinematic region in which the sum of large Pomeron loops can describe the scattering amplitude by estimating the contribution of the first small Pomeron loop in \fig{lpoml}-b.
 In QCD the vertices $\pom \to 2\,\pom \sim\,\bas$ and  $ 2 \,\pom \to \pom \,\sim\,\frac{\bas^3}{N^2_c}$ (see Ref.\cite{LELU}). These vertices lead to a contribution of the order of $\frac{\bas^4}{N^2_c} $  of the small Pomeron loop in  \fig{lpoml}-b. Therefore, we believe that in  QCD we can trust that the sum of large Pomeron loops will describe the scattering amplitude for $Y \,\leq\,\frac{N^2_c}{\bas^4}$ with a numerical factor in front.

~

~

\begin{boldmath}
\subsection{One dimensional model}
\end{boldmath}

We suggest considering  the one dimensional model to get  intuition for the answer to the  question: does the sum of large Pomeron loops satisfy  $s$-channel unitarity; and to find the kinematic region where we can trust the scattering amplitude that comes from summing  large Pomeron loops.

As we have mentioned, we still do not know the QCD Hamiltonian for Pomeron interaction that satisfy  $s$-channel unitarity.
 On the other hand such  a  Hamiltonian has been suggested \cite{KLL} for one dimensional models, which are very useful for getting intuitions about  what we need in QCD. It turns out that the Hamiltonian, proposed in Ref.\cite{KLL} leads to the one dimensional model,  suggested before in Ref.\cite{MUSA} based on quite different ideas: the scattering matrix {\cal S } does not depend on $Y_0$ (see \fig{lpoml}-a).
The total  $S$-matrix for scattering of $n$ dipoles of the projectile on $m$ dipoles of the target has the form
  \beq \label{SMS}
{\cal  S}(Y) \,\,=\,\,\sum_{n,m} e^{ - m \,n\,\gamma} \,P_n^P( Y_0)\,P_m^T( Y - Y_0)
 \eeq
where $n$ and $m$ are the number of dipoles in the projectile and target. $P_n$ are  the 
multiplicity distributions.

It turns out that the model\footnote{We call this model the  unitarity toy model: UTM, since it satisfies both $s$- and $t$- channel unitarity.}  of Ref.\cite{MUSA} is equivalent to the Hamiltonian  approach \cite{KLL}. It has been studied extensively during the past decade
\cite{MUSA,nestor,RS,SHXI,KOLEV,BMMSX,BIT,LEPRI,utm,utmm,utmp,LETMP}.
The scattering amplitude for dipole-dipole interaction has the following form \cite{utmp}
        for small $\gamma$:
\beq \label{SA}
{\cal S}\Lb \Y\Rb\,\,=\,\,
\,\,\,\,\sum^\infty_{n=0} \Lb - \gamma\Rb^n \,\,n! \,\,e^{ \Delta_n\,\Y}\,\,\Bigg\{ 1\,\,\,+\,\,{\cal O}\Lb \gamma\Rb\Bigg\}
\eeq 
$\Y = \frac{\Delta}{\gamma}\,Y$ where $ \Delta $ is the intercept of the Pomeron and $\gamma$ is the amplitude of the interaction of two dipoles at low energy. We will discuss this expression later. 

In one-dimensional models we do not have transverse coordinate integration and $\rho_n(Y - Y_0) = e^{n \Delta (Y -Y_0)}$. Plugging these $\rho_n$ into \eq{MPSI} we obtain that
\beq \label{ODM1}
{\cal S}\Lb Y\Rb\,\,=\,\,\sum_{n=0}^\infty (-1)^n\,n! \,\gamma^n\,\,e^{\Delta\, n \,Y}
\eeq
Comparing this equation with \eq{SA} one can see that the sum of large Pomeron loops reproduces the coefficients  in front of the Green's function for the exchange of $n$ Pomerons. However, the intercepts of these  functions are equal to $ \frac{\Delta}{\gamma} \Delta_n =\frac{\Delta}{\gamma} \Lb e^{\gamma\,n} - 1\Rb$ in the UTM, while for the sum of large Pomeron loops they are $\Delta\,n$. This does not look surprisingly, since the small Pomeron loops of  size $\delta Y \approx  \Delta$ contribute to the calculation of the intercepts,  as it  has been discussed in detail in Ref.\cite{LETMP} (see appendix B that paper). For $\gamma\,n\,\ll\,1$ $ \frac{\Delta}{\gamma} \Delta_n\,\,\to\,\Delta\,n$, but   this cannot be correct for large $n$. However, the situation looks quite differently for the behaviour of the amplitude at large $Y$. 

Both series of \eq{SA} and of \eq{ODM1} are asymptotic. The first one can be summed using the Borel procedure \cite{BORSUM}, while the second cannot be summed using this procedure. In Ref.\cite{utmp} a method of summing series of this kind has been developed. As a result we can rewrite both series as  integrals and
analytically continue then to the negative $n$. Indeed,
\beq \label{ODM2}
{\cal S}^{\rm UTM}\Lb \Y\Rb\,\,=\,\,
\,\,\,\,\int_C \frac{d n}{2\,\pi\,i} \frac{1}{\sin\Lb \pi\,n\Rb} \gamma^n \,\Gamma(n+1)\,\,e^{ \Delta_n\,\Y};~~~~~{\cal S}^{\rm LL}\Lb Y\Rb\,\,=\,\, \,\int_C \frac{d n}{2\,\pi\,i} \frac{1}{\sin\Lb \pi\,n\Rb} \Gamma(n+1)\,\gamma^n\,\,e^{\Delta\, n \,Y};
\eeq
where LL stands for large Pomeron loops.

     \begin{figure}[ht]
    \centering
  \leavevmode
      \includegraphics[width=6cm]{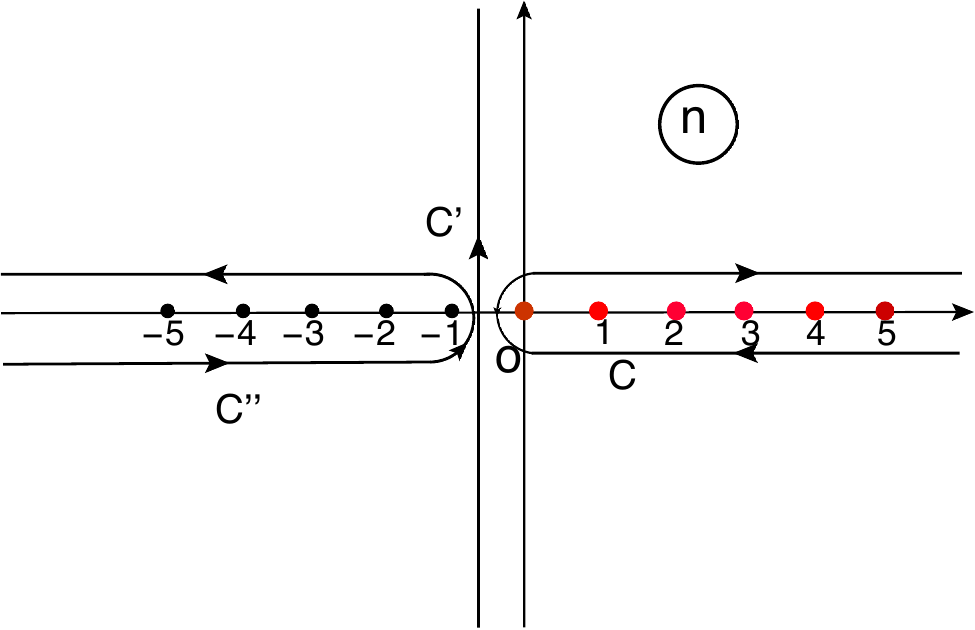}  
      \caption{The contours of integration in \eq{ODM2}. The red points denote the poles in $n$ for the right half-plane. The black points show the poles in the left half-plane where $S\Lb z\Rb$ has the double poles.}  
\label{cont}
   \end{figure}
After analytical continuation to  negative $n$ half-plane we can choose contour $C''$ and take the first double pole at $n=-1$,  which gives the asymptotic behaviour of the scattering amplitude at large values of $Y$. Hence
\beq \label{ODM20}
{\cal S}^{\rm UTM}\Lb \Y\Rb\,\,\xrightarrow{Y \,\gg\,1}\,\,\pi \ln\Lb \frac{\Delta}{\gamma} \Lb 1 - e^{-\gamma}\Rb Y\Rb \exp\Lb - \frac{\Delta}{\gamma} \Lb 1 - e^{-\gamma}\Rb\,Y\Rb;
~~~~~{\cal S}^{\rm LL}\Lb Y\Rb\,\,=\,\, \,\pi \ln\Lb \Delta \,Y\Rb \exp\Lb - \Delta\,Y\Rb;\eeq

One can see that for $Y\, \leq \,\frac{1}{\gamma^2}$ both expressions coincide.
In QCD $\gamma \sim \bas^2$, hence for the  large region of $Y \leq \frac{1}{\bas^4}$
we can claim that the sum of large Pomeron loops gives the scattering amplitude.

It is instructive to note that corrections to the BFKL intercept  due to the contributions of the  small Pomeron loops  are of the order of $\bas^4$ (see appendix B  of  Ref.\cite{LETMP}).

~

~

\begin{boldmath}
\subsection{Discussions: lessons for QCD}
\end{boldmath}

First, we can conclude that the calculation of  the contribution of the large Pomeron loops can be trusted for $Y\,\leq\,\frac{1}{\bas^4}$. These estimates come  from the first diagram of the small Pomeron loop of \fig{lpoml}-b and from the consideration of the one-dimensional model which satisfies both $t$- and $s$- channel unitarity.

Second, in the UTM we can investigate the fixed points of the equations of motion. The Hamiltonian of the UTM has the form \cite{KLL}:
\beq \label{DIS1}
H_{\mbox{\tiny UTM}}\Lb P, \bar P \Rb\,\,=\,\,-\,\frac{\Delta_{\tiny{BFKL}}}{\gamma} \bar P \,P
\eeq
with $\Phi = \ln\Lb 1 - P\Rb$ and $\bar \Phi =\ln \Lb 1 - \bar P\Rb$ being canonically conjugate. $\Delta_{\tiny{BFKL}}$ is the intercept of the BFKL Pomeron in one dimensional model. Below we absorb this intercept into the definition of  rapidities: $\tilde{Y} = \Delta_{\tiny{BFKL}}\,Y$ and $\tilde\eta = \Delta_{\tiny{BFKL}}\,\eta$.

The equations of motion take the form\cite{KLL}:
\beq \label{DIS2}
\frac{d\,P\Lb\tilde \eta\Rb}{d\,\tilde\eta} \,\,= \,\,\Lb 1\,-\,\bar P\Rb\,\Lb 1 \,-\,P\Rb\,P;~~~~~~~~~
\frac{d\,\bar P\Lb\tilde \eta\Rb}{d\,\tilde \eta}\,\, =\,\,-\, \Lb 1\,-\, P\Rb\,\Lb 1 \,-\,\bar P\Rb\,\bar P;
\eeq

At first sight these equations have the same fixed points as for $H_B$: (00),(1,0),(0,1) and (1,1). However the conservation of the Hamiltonian $\bar P \,P= {\rm Const}\Lb\eta\Rb\,\equiv \alpha$ completely changes  our view of the fixed points of the equations. They can be rewritten in the following form:
\beq \label{DIS3}
\frac{d\,P\Lb\tilde \eta\Rb}{d\tilde\eta} \,\,= \,\,\Lb  P\,-\,\alpha \Rb\,\Lb 1 \,-\,P\Rb;~~~~~~~~~
\frac{d\,\bar P\Lb\tilde\eta\Rb}{d\,\tilde\eta}\,\, =\,\,-\, \Lb \bar  P\,-\,\alpha\Rb\,\Lb 1 \,-\,\bar P\Rb\,;
\eeq
From these equations one can see that none   of  the  fixed points: (00),(1,0),(0,1), and (1,1), can be reached at $\alpha \neq 0$. We can see two fixed points in \eq{DIS3}: (1,$\alpha$) and ($\alpha, 1$).  Since $P(\eta=0)>0$ we can conclude that at $\eta \rightarrow\,\infty$ the asymptotics  is dominated by the fixed point (1,$\alpha$), while for $\eta \rightarrow \,0$ the main contribution comes from the fixed point ($\alpha$,1).

The solutions to \eq{DIS3} have been found in Ref.\cite{KLL} and they  have the form:
\beq \label{DIS31}
P\Lb \tilde \eta\Rb\,\,=\,\,1 - \frac{1- \alpha}{1 + \sqrt{\alpha} {\cal Z}\Lb\tilde Y,\tilde \eta\Rb};~~\bar P\Lb \tilde \eta\Rb\,\,=\,\,\alpha + \frac{(1- \alpha)\sqrt{\alpha}}{\sqrt{\alpha} + \ {\cal Z}\Lb\tilde Y,\tilde \eta\Rb};~~~ {\cal Z}\Lb\tilde Y,\tilde \eta\Rb
\,=\,e^{( 1 - \alpha) \Lb \tilde \eta - \h \tilde Y\Rb};
\eeq
For large $\eta$ and $Y$ \eq{DIS31} reduces to the following expressions:
\beq \label{DIS32}
P\Lb \tilde \eta = \tilde Y\Rb \xrightarrow{Y \gg 1}\, 1 - \frac{1-\alpha}{\sqrt{\alpha}} e^{ - \h(1- \alpha)\tilde Y} = 1 - \Delta\Lb \tilde Y\Rb;~~~~~~~
\bar P\Lb \tilde \eta = 0\Rb \xrightarrow{Y \gg 1}\,\alpha \,+\,(1-\alpha)\sqrt{\alpha}e^{ - \h(1- \alpha)\tilde Y} = \alpha + \bar \Delta\Lb \tilde Y\Rb;
\eeq

The scattering matrix {\cal S} for the UTM, which we have discussed in the previous section,  cannot be written as 
\beq \label{DIS4}
{\cal S}^{UTM}\,\,\sim\,\,(1 \,- \,P(\eta=Y))\,(1 \,-\,\bar P(\eta = 0)) \xrightarrow{\tilde Y \gg 1}(1 - \alpha) \Delta\Lb \tilde Y\Rb\eeq

Actually at $\gamma = \alpha \,\ll\,1$ this S-matrix has the form (see \eq{ODM2}):
\beq \label{DIS5}
{\cal S}^{UTM}\,\,\propto\,\, \Delta\Lb \tilde Y\Rb\,\bar \Delta\Lb \tilde Y\Rb
\eeq

Taking into account that the path integral of \eq{action2} takes the following form for the UTM\cite{KLL}:
\bea\label{DIS6}
&&{\cal S}^{UTM}(\Y)= \\
&&\int dP(\te)d\bar P(\te)e^{S_0+S_I}(1-P(\Y))(1-\bar P(0 )) =
\int dP(\te)d\bar P(\te)e^{\frac{1}{\gamma} \int_0^Yd\te\left[ \ln(1-P)\frac{\partial}{\partial \te}\ln (1-\bar P) 
+\bar PP\right]}(1-P(\Y))(1-\bar P(0))\nn\\
&&=\int dP(\te)d\bar P(\te)e^{\frac{1}{\gamma}\Lb\ln(1-P)\ln\Lb 1-\bar P\Rb\Big{|}^{\te =\Y}_{\te= 0} \,-\, \int_0^Yd\te\left[ \frac{\partial}{\partial \te}\ln (1- P)\ln(1-\bar P) 
+\bar PP\right]\Rb}(1-P(\Y))(1-\bar P(0))\nn\\
&&=\int dP(\te)d\bar P(\te)e^{\frac{1}{\gamma}\Lb\,-\, \int_0^{\Y}d\te\left[ \frac{\partial}{\partial \eta}\ln (1- P)\ln(1-\bar P) 
+\bar PP\right]\Rb}(1-P(\Y))^2\,\xrightarrow{Y\,\gg\,1}(1-P(\Y))^2 = \Delta^2 \Lb \tilde Y\Rb\nn
\eea
One can see that explicit calculations with the solutions of \eq{DIS31} reproduces the result of the exact solution which has been discussed in the previous subsection.

Hence, the contribution of $S_0+S_I$ in the path integral for the solutions of the equation of motion is essential  (cf. section III, \eq{action22}). It is worthwhile mentioning that the resulting solution leads to the asymptotic behaviour of ${\cal S}^{UTM}$ 
which would be expected in the case of the fixed point (1,1). 

Concluding, we wish to point out that :  (i) the UTM shows that summing the large Pomeron loops leads to the correct scattering amplitude for large range of rapidities: $ Y\,\,\leq\,\,1/\gamma^2$, (ii) for  $Y \,>\,1/\gamma^2$ the small Pomeron loops with size of about $ \delta Y\, \sim\, \Delta_{\tiny{BFKL}}$ become important and they change drastically the intercepts of multi-Pomeron exchanges; and (iii) \eq{action3} is not correct but the  correct asymptotic behaviour ${\cal S} \sim\,\Delta^2(Y)$ coincides with
\eq{action3}  if the Hamiltonian had a fixed point (1,1).

~

~

\begin{boldmath}
\section{Conclusions}
\end{boldmath}

\subsubsection{The main results}

  ~
  
  The main result of this paper is \eq{I0} for the scattering amplitude in the Pomeron field approach to the high energy QCD. This result is based on 
 the Braun Hamiltonian for the Pomeron interaction, which holds in QCD
at a large number of colours ($N_c$).   
   It is worthwhile mentioning that we obtain \eq{I0}  in the same theoretical framework as it was done for the  BK cascade in Ref.\cite{LETU}.  We have demonstrated this point in \eq{2AEQ5}.
  
  However, it should be noted that both the  Braun and BK Hamiltonians have problems with $s$-channel unitarity\cite{KLLL,KLLL1}. It is important to emphasize that the origin of this discrepancy is the same for both Hamiltonians. 
Heeling  these problems,  we demonstrated 
in the one-dimensional model (UTM), which satisfies both $t$- and $s$- channel unitarity
that the approach, based on \eq{action3},  could be inconsistent. Indeed, in the UTM we found the equations of motion(see \eq{DIS2}) with the same fixed points as in the  Braun hamiltonian approach: (0,0),{1,0},(0,1) and (1,1). As it has been discussed,  the conservation of energy reduces these points to two fixed points:(1,$\alpha$) and ($\alpha$,1) (see \eq{DIS3}). 
However, the final answer for the scattering amplitude in the UTM leads to 
 ${\cal S} \sim\,\Delta^2(Y)$  which coincides with
\eq{action3}  if the Hamiltonian had a fixed point (1,1). Therefore, we have  mixed feelings. On one hand, we are pessimistic since the form of the Hamiltonian can change crucially the structure of the fixed points for the equations of motion. Recall, that our solution is based on this structure. On the other hand, we can feel rather optimistically, since the scattering amplitude for the USM could be found using \eq{action3}.
  
  The large $z$ behaviour of the dipole-dipole scattering amplitude given by \eq{I0} or \eq{I01} is in  great  discrepancy both with 'rare' fluctuations\cite{IAMU,IAMU1} and with the procedure of summing large Pomeron loops\cite{LELT,LEDIDI}. 
  
\subsubsection{'Rare' fluctuations versus our results}
  
We do not have any explanation why `rare` fluctuations do not reproduce our theoretical result. However, our result is based on the assumption that the main contribution to the behaviour of the scattering matrix in the saturation region comes from $ \Phi  \to 1$ and $\bar{\Phi}\,\to\,1$.  We did not see any violation of  $s$ and $t$ unitarity in this region. However,  we cannot prove that other regions lead to  smaller contributions. For example,  the region where $\Phi \,\to\,1$ and $\bar{\Phi}$ is small, gives the larger  value of S-matrix but it faces a problem with $s$-channel unitarity. It should be stressed that this particular region would be preferred from the point of view of `rare' fluctuation approach, which evaluate a contribution only from the point of view of its influence on the value of  the scattering amplitude.
We also wish to point out that we obtain the solution for the scattering amplitude in this kinematic region which coincides with the solution to the BK equation\cite{LETU}. This is in  a clear contradiction with the statement of Ref.\cite{IAMU} that the real asymptotic behaviour has to be determined by the equation with a different shadow correction term than in BK equation.

As far as summing large Pomeron loops, we demonstrated in this paper that we can trust their contributions to the scattering amplitude only for $Y \,\leq\,1/\as^4$ with the numerical factor in front which we failed to estimate. It should be mentioned that summing the large Pomeron loops leads to the same result as the `rare' fluctuation approach. We are going the study this in our future publications. 

It is worth mentioning that  the UTM shows that summing the large Pomeron loops leads to the scattering amplitude which differs from exact asymptotic expression by the intercept of the Pomeron. We need to sum short Pomeron loops to obtain the correct value of the intercept. As we have seen, the solution in high energy QCD cannot be reduced to the change of the Pomeron intercept. 

Frankly speaking, the 'rare' fluctuation ideas call for more microscopic approach than the BFKL Pomeron calculus. Therefore, it is instructive to collect what we know about this calculus. First, as we have mentioned in the introduction, the Pomeron calculus describes the Color Glass Condensate (CGC) approach\cite{JIMWLK1,JIMWLK2, JIMWLK3,JIMWLK4, JIMWLK5,JIMWLK6,JIMWLK7,JIMWLK8} (see also Ref.\cite{KOLEB} for a review). Second, the BFKL Pomeron calculus leads to the scattering amplitude that satisfies the t-channel unitarity, due to the reggeization of gluons.
Third, the exact solution to the scattering amplitude in the one dimensional (UTM) model can be derived in the Pomeron calculus as well as in the microscopic approach, which is based on dipole interactions\cite{KLLN}. Bearing this in mind we believe that the BFKL Pomeron calculus can be a reliable approach for high energy QCD in spite of  problems with s-channel unitarity.

 This paper is a direct continuation of our old paper of Ref.\cite{KLL}. We firmly believe that both papers  call for more effort in finding  the QCD Hamiltonian for Pomeron interactions which satisfies both $t$- and $s$- channel unitarity. It looks unlikely that we can find the scattering amplitude at large $z$ without such a Hamiltonian even for Balitsky-Kovchegov scattering amplitude.
 
~

~

   {\bf Acknowledgements} 
     
   We thank our colleagues at Tel Aviv University  for
 discussions. Special thanks go A. Kovner and M. Lublinsky for being coauthors of Refs.\cite{AKLL,ACKLLS,KLL,KLLL1,KLLL,KLLN} and for
 stimulating and encouraging discussions on the subject of this paper. 
  This research was supported  by 
BSF grant 2022132.

\end{document}